# Biomimetic method for metallic nanostructured mesoscopic models fabrication

**Gennady V. Strukov and Galina K. Strukova**

Institute of Solid State Physics, Russian Academy of Sciences, 142432 Chernogolovka, Russia

*E-mail: strukov@issp.ac.ru

Scientists have long been trying to understand the laws of biological growth and morphogenesis whereas material scientists and engineers have been making models of wonderful nature creations to reproduce their useful functional properties. The fact is that millions of years of evolution have created materials with characteristics that are still unattainable for artificial materials. The phenomenal self cleaning ability of a lotus leaf and unique mechanical stiffness of a nacre may serve as good examples. A number of papers are devoted to studies of the structure of natural shells of molluscan shellfish exhibiting the mechanical properties that are so far unfeasible for synthesized materials and to development of methods of their model synthesis. Among the unsolved problems of biomimetics is that of reproducing the structure of nacre exhibiting a phenomenal external load stability. There is an intensive search for materials capable of copying natural ones and methods of synthesizing them, i.e., "biomimetic" methods (as an example [1,2] ). It was found, that the superior properties of natural materials are due to their architecture and a hierarchical structure. The creation of hierarchical structure of man-made material in their synthesis process is a crucial task for biomimetic method. Another important issue: is shape control of model and their reproducible fabrication feasible?

At the Institute of Solid State Physics RAS where materials science is one of the major research fields, we studied nanostructured metallic coatings and nanowires with magnetic and superconducting properties. In the course of nanowire growth by electrodeposition of metal on porous membranes our attention was attracted by the meso-structures forming on the membrane. These structures grow on porous membranes by means of pulsed current electrodeposition if the electroplating is continued after the nanowires appear on the membrane surface. There are two scenarios possible for growing metal nanowires in porous membranes by means of pulse current electrodeposition. The first one is when nanowires, after they have appeared on the membrane surface, continue to grow separately. The second one, considered in the present paper, is when nanowires form nanostructured "vegetable" metallic meso-samples on the membrane surface. This occurs in the case of self-assembly of nanowires that appear on the membrane surface simultaneously with relatively small distances between the nanowires. Both methods were implemented by us.

The results presented show that pulse current electrodeposition on porous membranes ensures controlled growth of nanostructured metallic models of natural objects, plants and fungi, i.e., it is a biomimetic method of their synthesis. We have succeeded in preparing pores of definite configuration on a polymer membrane, which together with precisely fixed regime of pulsed current electroplating, results in single type convex-concave structures resembling shells.

The architecture of the "shells" was revealed by their fragmentation in an ultrasound bath and chemical etching. The self-organisation of the metallic nanowires grown from the membrane gave rise to nanosize conical elements acting as "bricks" to build a layered hierarchic structure at the nano-, micro- and meso-levels. Our models appear to replicate not only the exterior form of biological objects but also their hierarchic structure.

Structural diversity, controllable shaping and the remarkable resemblance to mushrooms, plants, shells as well as the hierarchic structure suggest that pulse current electrodeposition on templates can be regarded as a biomimetic method of creating metallic structures. These arguments allow us to put forward the hypotheses that pulsed growth on templates is a tool of morphogenesis for most mushrooms and plants which is accompanied by self-organisation of growing clusters and fibers and fractal branching. By varying pore size and pore pattern in a membrane, and by varying the electrolytes and the pulsed current parameters, an impressive manifold of metallic nanostructured mesostructures resembling natural objects (mushrooms and plants) was produced [3,4.5]. Nanostructured models have been grown from normal (Ag, Pd, Rh), magnetic (Ni, Pd-Ni, Pd-Co) and superconducting (Bi, Pb-In, Pb-Bi) metals and alloys which opens up prospects for their use in creation of nanodevices. Nanostructured large surface models such as "silver wood" and similar structures from other metals are interesting for applications such as catalytic filters, batteries and supercapacitors.

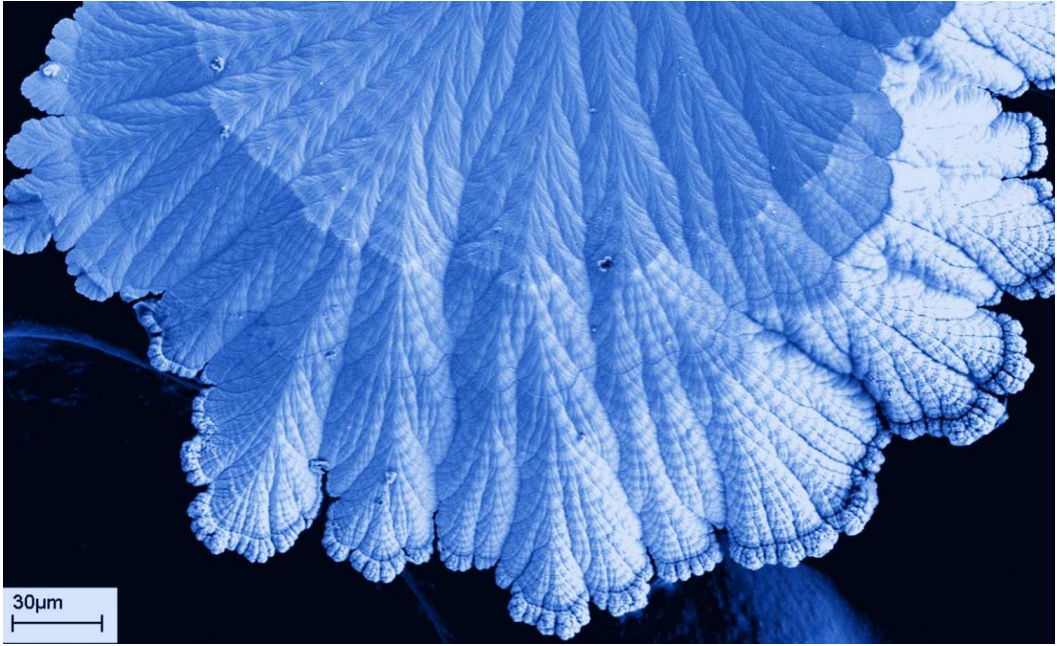

Fig. 1

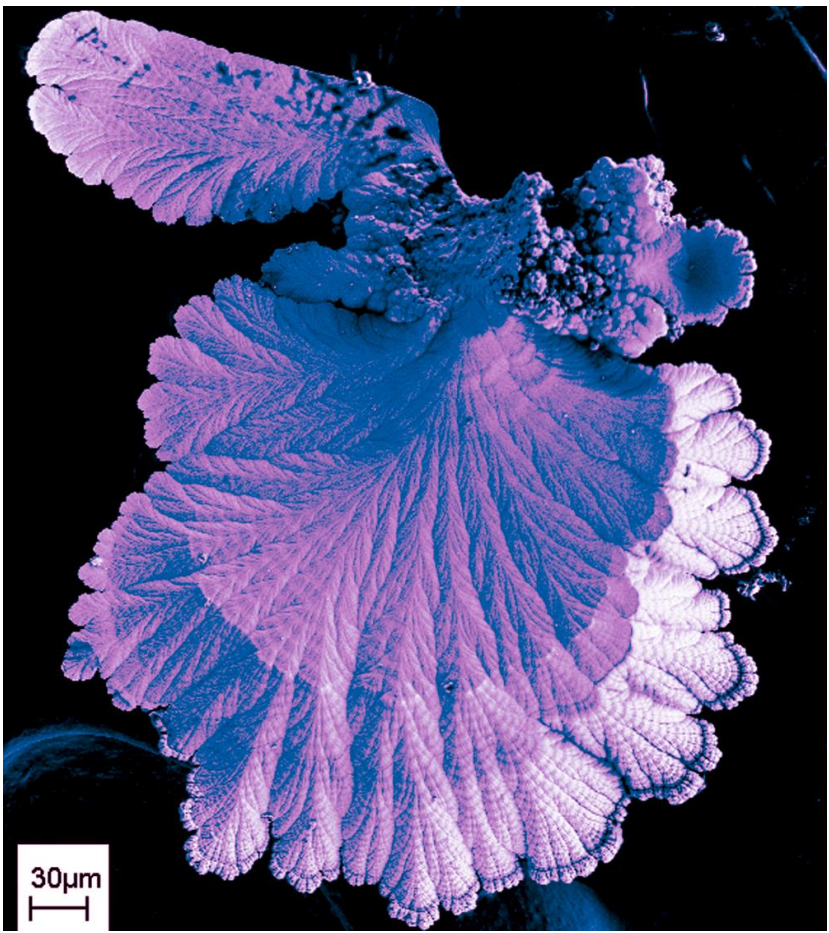

Fig.2

This fig. 1 and fig.2 images shows mesostructure of Pd-Ni-alloy grown from specifically prepared pores. The image is made by using a scanning electron microscope SUPRA -50 VP. These delicate leaves are reproduced as a result of self-organisation of nanowires growing on porous membrane in the course of pulse current electrodeposition. "Leaves" start growing from the "root" ("bottom") which is a stub -like site of layered fibers. The grown "leaves" are example of metallic woven multilayer surface. The inner "leaf" surface exhibits a nonuniform relief, a pronounced woven pattern, its lines directed from the "root" to the periphery. This relief is a manifestation of the inner architecture on the surface layer. Due to its technological simplicity, template growth of nanostructured metallic coatings may be used to fabricate superhydrophobic surfaces for technical applications.